\documentclass[12pt,showpacs,preprintnumbers,amssymb]{iopart}
\usepackage{graphicx}
\usepackage{float}
\usepackage{verbatim}
\usepackage{dcolumn}
\usepackage{bm}

\begin{document}


\newcommand{\beq}{\begin{equation}}
\newcommand{\eeq}{\end{equation}}
\newcommand{\beqa}{\begin{eqnarray}}
\newcommand{\eeqa}{\end{eqnarray}}
\newcommand{\lf}{\hfil \break \break}
\newcommand{\ahat}{\hat{a}}
\newcommand{\adag}{\hat{a}^{\dagger}}
\newcommand{\adagg}{\hat{a}_g^{\dagger}}
\newcommand{\bhat}{\hat{b}}
\newcommand{\bdag}{\hat{b}^{\dagger}}
\newcommand{\bdagg}{\hat{b}_g^{\dagger}}
\newcommand{\chat}{\hat{c}}
\newcommand{\cdag}{\hat{c}^{\dagger}}
\newcommand{\dhat}{\hat{d}}
\newcommand{\nhat}{\hat{n}}
\newcommand{\ndag}{\hat{n}^{\dagger}}
\newcommand{\den}{\hat{\rho}}
\newcommand{\phihat}{\hat{\phi}}
\newcommand{\Ahat}{\hat{A}}
\newcommand{\Adag}{\hat{A}^{\dagger}}
\newcommand{\Bhat}{\hat{B}}
\newcommand{\Bdag}{\hat{B}^{\dagger}}
\newcommand{\Chat}{\hat{C}}
\newcommand{\Dhat}{\hat{D}}
\newcommand{\Ehat}{\hat{E}}
\newcommand{\Lhat}{\hat{L}}
\newcommand{\Nhat}{\hat{N}}
\newcommand{\Ohat}{\hat{O}}
\newcommand{\Odag}{\hat{O}^{\dagger}}
\newcommand{\Shat}{\hat{S}}
\newcommand{\Uhat}{\hat{U}}
\newcommand{\Udag}{\hat{U}^{\dagger}}
\newcommand{\Xhat}{\hat{X}}
\newcommand{\Zhat}{\hat{Z}}
\newcommand{\Xdag}{\hat{X}^{\dagger}}
\newcommand{\Ydag}{\hat{Y}^{\dagger}}
\newcommand{\Zdag}{\hat{Z}^{\dagger}}
\newcommand{\Ham}{\hat{H}}
\newcommand{\bis}{{\prime \prime}}
\newcommand{\tris}{{\prime \prime \prime}}
\newcommand{\ket}[1]{\mbox{$|#1\rangle$}}
\newcommand{\bra}[1]{\mbox{$\langle#1|$}}
\newcommand{\ketbra}[2]{\mbox{$|#1\rangle \langle#2|$}}
\newcommand{\braket}[2]{\mbox{$\langle#1|#2\rangle$}}
\newcommand{\bracket}[3]{\mbox{$\langle#1|#2|#3\rangle$}}
\newcommand{\dotp}{\mbox{\boldmath $\cdot$}}
\newcommand{\tp}{\otimes}
\newcommand{\op}[2]{\mbox{$|#1\rangle\langle#2|$}}
\newcommand{\hak}[1]{\left[ #1 \right]}
\newcommand{\vin}[1]{\langle #1 \rangle}
\newcommand{\abs}[1]{\left| #1 \right|}
\newcommand{\tes}[1]{\left( #1 \right)}
\newcommand{\braces}[1]{\left\{ #1 \right\}}
\newcommand{\indist}{I}
\newcommand{\initial}[2]{|\psi^#1_#2(0)\rangle}
\newcommand{\initialbra}[2]{\langle\psi^#1_#2(0)}
\newcommand{\decohered}[2]{|\psi^#1_#2(\gamma)\rangle}
\newcommand{\decoheredbra}[2]{\langle\psi^#1_#2(\gamma)}
\newcommand{\gendecohered}[1]{|\psi^G(#1)\rangle}
\newcommand{\gendecoheredbra}[1]{\langle\psi^G(#1)}
\newcommand{\projector}[2]{|\ksi^#1_#2 \rangle}


\title[Non-monotonic projection probabilites]{Non-monotonic dependence of projection probabilities as a function of distinguishability}
\author{Gunnar Bj\"{o}rk, Saroosh Shabbir}
\address{Department of Applied Physics, Royal Institute of Technology (KTH)\\
AlbaNova University Center, SE - 106 91 Stockholm, Sweden}
\ead{saroosh@kth.se}

\date{\today}

\begin{abstract}
Typically, quantum superpositions, and thus measurement projections of quantum states involving interference, decrease (or increase) monotonically as a function of increased distinguishability. Distinguishability, in turn, can be a consequence of decoherence, for example caused by the (simultaneous) loss of excitation or due to inadequate mode matching (either deliberate or indeliberate). It is known that for some cases of multi-photon interference, non-monotonic decay of projection probabilities occurs, which has so far been attributed to interference between four or more two photons. We show that such a non-monotonic behaviour of projection probabilities is not unnatural, and can also occur for single-photon and  even semiclassical states. Thus, while the effect traces its roots from indistinguishability and thus interference, the states for which this can be observed do not need to have particular quantum features.
\end{abstract}

\maketitle

\section{Introduction}
Interference is the manifestation of the quantum mechanical law stating that when a final state can be reached through two or more indistinguishable ``paths'', the path probability amplitudes should be added before any final state probability is calculated. Since probability amplitudes are represented by complex numbers and may change signs, path probability amplitudes can thus ``annihilate'' each other, resulting in destructive interference. On the contrary, if the paths to the final state are distinguishable, if even only in principle, then the probabilities and not the amplitudes of the paths should be added. Since probabilities are real and non-negative, adding probabilities cannot result in destructive interference.  Perhaps the best known experiment to demonstrate the general validity of this law, i.e., interference due to indistinguishability is the Hanbury Brown and Twiss stellar interferometer \cite{HBT,HBT2}. The surprise for the scientific community at the time was that intensities interfered, rather than wave amplitudes. However, this experiment firmly established that path indistinguishability is what causes interference.\\

Thus, it is perhaps to be expected that constructive (destructive) interference probabilities should decrease (increase) monotonically with increasing distinguishability. In either case one could consider that when interference is probed as a function of distinguishability a monotonic behaviour would follow. The Hong-Ou-Mandel (HOM) experiment \cite{HOM} is a prototype for examining the interference of increasingly distinguishable particles, which indeed shows monotonic dependence as a function of distinguishability. However, recent extensions of the HOM interference to higher photon numbers \cite{Ra, Tichy} have shown that under certain circumstances the measured interference signal varies non-monotonically with increasing distinguishability of the interfering particles (photons in this case).\\

The quantitative relationship between path distinguishability and interference visibility has been studied in the past in different contexts \cite{Jaeger,Englert,Durr,Bjork,Gavenda}. Here, we are not going to concern ourselves with visibility, but only look at how certain (projective) measurements of interference depend on the degree of distinguishability, and specifically whether they vary monotonically with increasing distinguishability or not. We thus do not study interference as a function of the relative phase between the paths, but for a fixed phase. Moreover, while first order interference oscillates sinusoidally with a path relative-phase period of $2 \pi$, the higher (even) order interference functions we will study (like the Hong-Ou-Mandel dip) do not oscillate as a function of the path relative-phase.\\

The authors of Ref. \cite{Ra, Tichy} associate distinguishable events with classical ``particles'' (or ``paths'') and the indistinguishable ``particles'' with quantum trajectories. They subsequently attribute the non-monotonicity to a quantum-to-classical transition of multi-particle interference. However, our analysis shows that contrary to what one may believe, interference probabilities can be non-monotonic for single particles as well, whether they are bosons or fermions. Moreover, the measured interference probabilities analysed and measured in \cite{Ra} were sums of, in principle distinguishable, final events, some increasing and some decreasing with increasing distinguishability. Taking the analysis further, we shall demonstrate that more fundamentally, isolated probabilities (projection probabilities onto pure, final states) can also vary non-monotonically. Finally we shall discuss that this behaviour occurs also classically, specifically with a classical light-source and both with conventional photo detectors and single-photon detectors. Thus we find that the non-monotonic behaviour is natural for all interfering particles or fields and is thus, in general, not a manifestation of a classical-to-quantum transition, or ``exclusive to setups with four or more particles that interfere simultaneously" \cite{Ra}.

\section{Analysis of the Hong-Ou-Mandel experiment} 
\begin{flushleft}
\begin{figure}[H]
\centering
\includegraphics[scale=0.9]{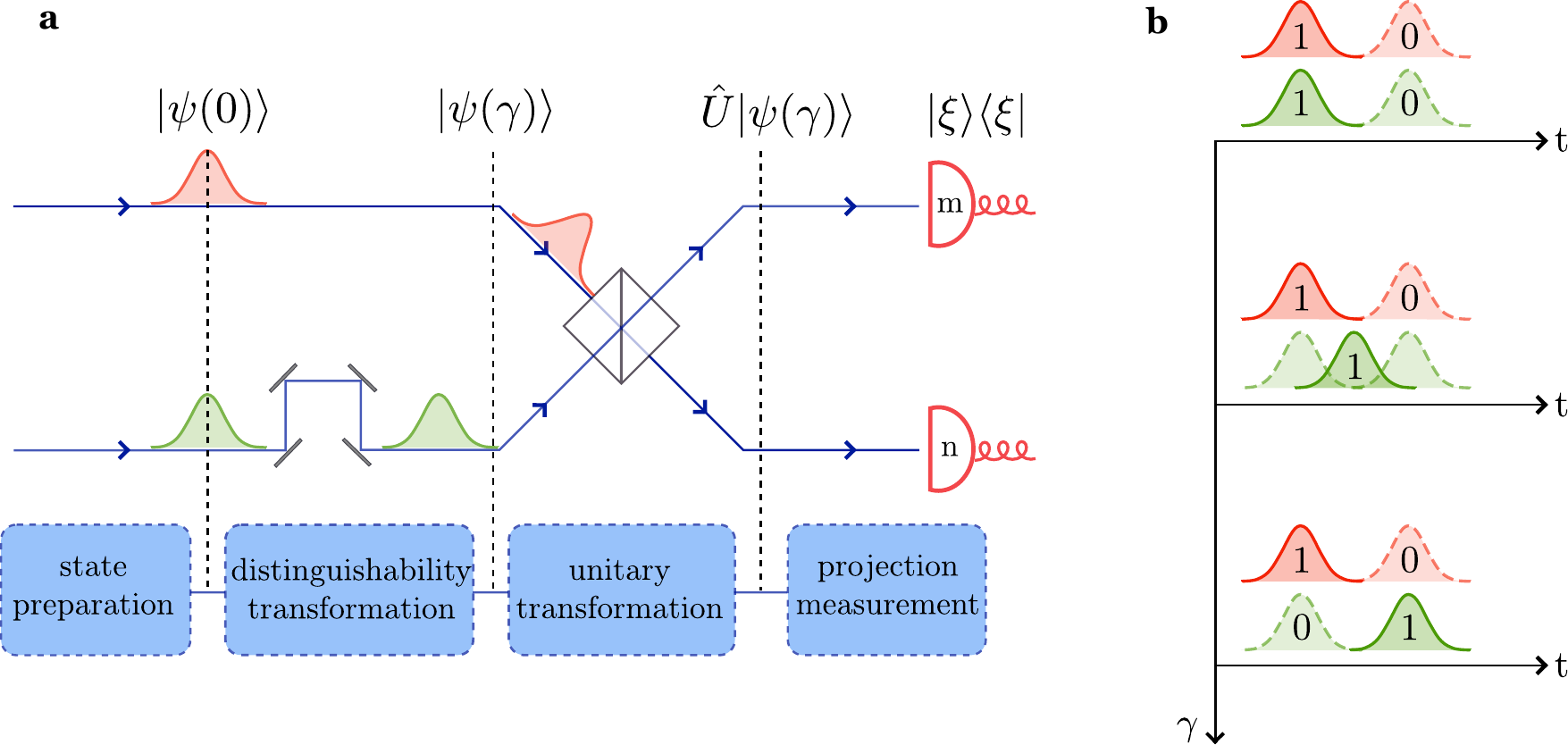}
\fl \caption{\textbf{a.} Schematic of the Hong-Ou-Mandel experiment. \textbf{b.} Early (left) and late (right) photon modes for each degree of freedom (red and green). Note that the t-axis represents the time at which the photons impinge on the beam splitter. $\gamma$ is the distinguishability transformation which increases downwards, so the top (bottom) figure represents the case when the temporal modes are perfectly indistinguishable (distinguishable).}
\label{Fig: schematic}
\end{figure}
\end{flushleft}

The Hong-Ou-Mandel (HOM) dip \cite{HOM} demonstrates how projection probabilities change as a function of distinguishability, where distinguishability is introduced by delaying one input photon with respect to the other in the second input port of a beam splitter (Fig.~\ref{Fig: schematic}a). The delay causes the loss of interference of the photons at the beam splitter, since time resolution of the output can in principle convey ``which-path" information.

\newpage

We model the HOM experiment with the four mode input state \beq \decohered{{\scriptsize \textrm{HOM}}}{2} = \cos(\gamma)\ket{1,1}\otimes \ket{0,0} + \sin(\gamma)\ket{1,0}\otimes \ket{0,1},\eeq where $\gamma$ parameterizes the delay, and hence distinguishability, between the two photons and ranges from $0 \leq \gamma \leq \pi/2$ and the subscript 2 indicates that it is a two-photon state. The two states to the left of the tensor multiplication sign refers to early photon modes, indistinguishable except for one degree of freedom such as path (illustrated in Fig.~\ref{Fig: schematic}b) or polarization, whereas the two states to the right refer to late photon modes, perfectly distinguishable to the early photons, but indistinguishable among themselves except for, e.g., the path. In this case, the time of arrival can fully determine which photon impinges on the beam splitter first, hence the designation early and late for the two photon modes. Thus when $\gamma=0$ the input photons are perfectly indistinguishable and can display maximal interference when the two modes interact in a 50:50 beam splitter described by the unitary operator $\hat{U}$, \emph{viz.}, $\hat{U}\ket{1,1}\otimes \ket{0,0} = (\ket{2,0}-\ket{0,2})\otimes \ket{0,0}/\sqrt{2}$.\\

Qualitatively, the indistinguishability of the input state can be defined as its projection probability onto the maximally interfering state, i.e., $\indist \equiv | \initialbra{{\scriptsize \textrm{HOM}}}{2}\decohered{{\scriptsize \textrm{HOM}}}{2}|^2=\cos^2(\gamma).$  Note that $\indist$ is not a good measure in quantitative terms: for different perfectly distinguishable modes $\indist$ may not be zero as we shall see below. However, for our purposes this simple measure suffices as it is a monotonically decreasing function of $\gamma$ for all our investigated methods of increasing distinguishability.

The output state after the beam splitter is
\begin{eqnarray}\fl \hat{U} \decohered{{\scriptsize \textrm{HOM}}}{2}=\frac{\cos(\gamma)}{\sqrt{2}}(\ket{2,0}-\ket{0,2})\otimes \ket{0,0} \nonumber \\ + \frac{\sin(\gamma)}{2}(\ket{1,0}+\ket{0,1})\otimes (\ket{1,0}-\ket{0,1}).\end{eqnarray} Unfortunately, it is rather difficult to measure the projection onto the state $\hat{U}\initial{{\scriptsize \textrm{HOM}}}{2}$. The remedy in a typical experiment is to instead project onto $\ket{1,0,0,1}\bra{1,0,0,1}$ and thus instead measure the \textit{distinguishability}. The state $\ket{1,0,0,1}$ represent the case when an early photon takes the first ``path'' and a late photon takes the second ``path''. If the photons are measured in coincidence, but the coherence length and the time delay of the modes are much shorter than the coincidence measurement window, they will still be recorded as coincident. This is a typical experimental situation. In this case the state $\ket{0,1,1,0}$ will also be detected as coincident, and although the two states are orthogonal they will both contribute to this measurement of interference. Mathematically, since the events are distinguishable in practice (by, e.g., noting in which mode the early photon arrived) the respective measurement probabilities should be added. Then, the output projection probability can be written as \beq \fl |\bra{1,0,0,1}\hat{U} \decohered{{\scriptsize \textrm{HOM}}}{2}|^2 + |\bra{0,1,1,0}\hat{U} \decohered{{\scriptsize \textrm{HOM}}}{2}|^2 = \frac{\sin^2(\gamma)}{2} \nonumber  = \frac{1-\indist}{2}. \eeq Thus, if $\indist$ is monotonic, so is this measurement of interference.\\

More generally, consider a state $\gendecohered{\gamma}$ of one or more photons, where $\gamma$ parameterizes the degree of distinguishability. For $\gamma=0$ the input modes are perfectly indistinguishable (except for one degree of freedom) when they interfere in some device described by the unitary transformation $\hat{U}$. If the input modes are made partially distinguishable, e.g., by shifting the modes in time, in polarization or in frequency, so that a state with $\gamma \neq 0$ is prepared, then the indistinguishability can be defined as the states' overlap with the maximally interfering state, i.e., $\indist_{\scriptsize \textrm{HOM}} \equiv |\gendecoheredbra{0}\gendecohered{\gamma}|^2$. After the unitary (interference) transformation, it is very natural to define the degree of indistinguishability as the output state's projection probability onto the output state state exhibiting maximal interference, i.e., $\hat{U}\gendecohered{0}$. In the following we shall define $\hat{U}\gendecohered{0} \gendecoheredbra{0}|\hat{U}^\dagger$  as the \textit{proper} indistinguishability projector.\\

Hence, independent of the exact nature of the interference between the photons, the degree of indistinguishability at the output will be $\indist_{\scriptsize \textrm{out}} = |\gendecoheredbra{0}|\hat{U}^\dagger \hat{U}\gendecohered{\gamma}|^2 = \indist_{\scriptsize \textrm{in}} $. From this, almost trivial, consideration we conclude that if the indistinguishability of the input varies in a monotonic fashion, then the \textit{properly measured} interference, and thus indistinguishability of the output modes, also varies monotonically. However, if one chooses a different measurement to probe the interference propensity, then one may obtain a non-monotonic probability as a function of distinguishability, e.g., by taking a sum of projections such that one increases with distinguishability while the other decreases. We demonstrate this in the two photon-pair HOM experiment analyzed below.\\

\section{Two photon-pair HOM experiment} The case where a pair of photons is delayed with respect to another pair at the input of the beam splitter \cite{Ra,Tichy} can be similarly modelled as \begin{eqnarray} \fl \decohered{{\scriptsize \textrm{HOM}}}{4} = \cos^2(\gamma)\ket{2,2}\otimes \ket{0,0} + \sqrt{2}\cos(\gamma)\sin(\gamma)\ket{2,1}\otimes \ket{0,1} \nonumber \\+ \sin^2(\gamma)\ket{2,0}\otimes \ket{0,2}.\end{eqnarray}  Using the ``recipe'' above, the indistinguishability between the two pairs of photons can be quantified by $\indist = \cos^4(\gamma)$. The indistinguishability thus decreases monotonically with $\gamma$ in the relevant interval $0\leq \gamma\leq \pi/2$. In the case of two photons in each output, the coincidence detection window can again be set so that the photons in the output states $\ket{2,2,0,0}$, $\ket{2,1,0,1}$, $\ket{1,2,1,0}$ $\ket{2,0,0,2}$, $\ket{1,1,1,1}$, and $\ket{0,2,2,0}$ all are defined to be coincident (i.e. two photons are detected in each of the two spatial or polarization modes within the coincidence window.) The individual probabilities for these projections are as follows:

\newpage

\begin{eqnarray}
  |\bra{2,2,0,0}\hat{U} \ket{\xi(\gamma)} |^2 &=& \frac{\cos^4(\gamma)}{4}, \nonumber \\
  |\bra{2,1,0,1}\hat{U} \ket{\xi(\gamma)} |^2 &=& \frac{\cos^2(\gamma)\sin^2(\gamma)}{8}, \nonumber \\
  |\bra{1,2,1,0}\hat{U} \ket{\xi(\gamma)} |^2 &=& \frac{\cos^2(\gamma)\sin^2(\gamma)}{8}, \label{Eq: projs 2 pair}\\
  |\bra{2,0,0,2}\hat{U} \ket{\xi(\gamma)} |^2 &=& \frac{\sin^4(\gamma)}{16}, \nonumber \\
  |\bra{1,1,1,1}\hat{U} \ket{\xi(\gamma)} |^2 &=& \frac{\sin^4(\gamma)}{4}, \nonumber \\
  |\bra{0,2,2,0}\hat{U} \ket{\xi(\gamma)} |^2 &=& \frac{\sin^4(\gamma)}{16}. \nonumber
\end{eqnarray}

\begin{figure}[h!]
\centering
\includegraphics[scale=0.5]{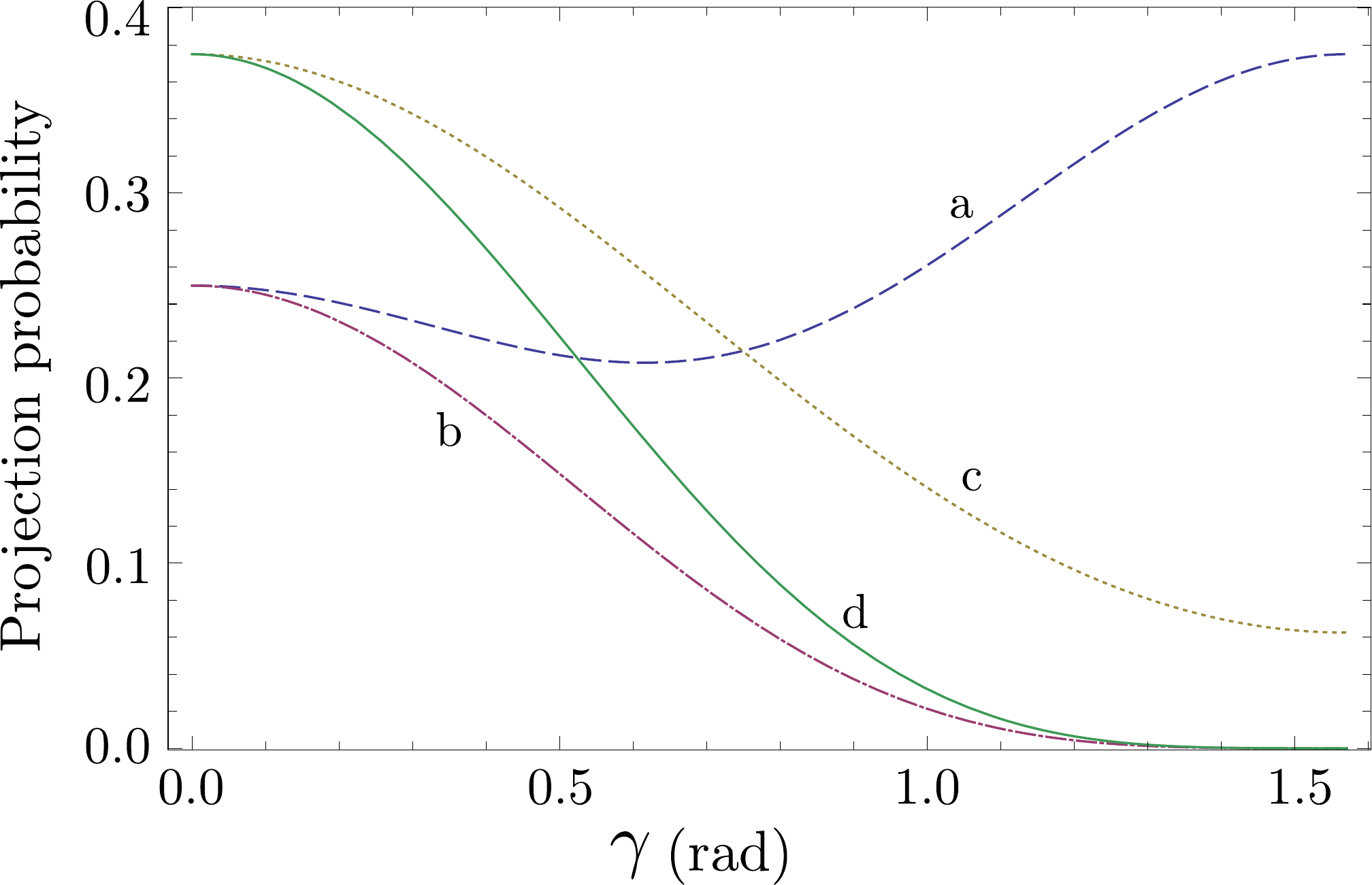}
\caption{Probabilities of the $(m,n,\tilde{m},\tilde{n})$ events for the two-pair input HOM experiment, where the latter two digits with tilde refer to late output modes of the beam splitter. \textbf{a.} Sum of (2,2,0,0), (2,1,0,1), (1,2,1,0), (2,0,0,2), (1,1,1,1) and (0,2,2,0) event probabilities. \textbf{b.}  (2,2,0,0) event probability. \textbf{c.} Sum of (4,0,0,0), (3,0,1,0) and (2,0,2,0) event probabilities. \textbf{d.} (4,0,0,0) event probability.}
\label{Fig: projection probabilities}
\end{figure}

We see that only the first of these projection probabilities is proportional to the indistinguishability of the input state. Collecting the terms contributing to the coincidence signal (under the assumption that the detection coincidence window $>$ the maximal time delay between the modes $>$ the coherence time) we get the coincidence count probability \beq P_4^{\scriptsize \textrm{HOM}} = \frac{\cos^4(\gamma)}{4} + \frac{\cos^2(\gamma)\sin^2(\gamma)}{4} + \frac{3 \sin^4(\gamma)}{8}.\eeq The probability in this case is not monotonic with respect to the distinguishability parameter $\gamma$ as can be seen in Fig. \ref{Fig: projection probabilities} and was experimentally verified in \cite{Ra}. To see this more clearly, the equation can be rewritten as \beq P_4^{\scriptsize \textrm{HOM}} = \frac{1}{8}\left (3 \indist -4 \sqrt{\indist} + 3 \right ).\eeq The function is non-monotonic partially because it measures a sum of terms, three growing with increasing distinguishability and one decreasing. Moreover, the second and third probabilities in (\ref{Eq: projs 2 pair}) are non-monotonic by themselves. For the bunching event (e.g., when all photons exit in the lower path or in the same polarization mode), the measured coincident probability is the sum of projections on $\ket{4,0,0,0}$, $\ket{3,0,1,0}$, $\ket{2,0,2,0}$ states. However, in this case the coincidence probability fortuitously turns out to be monotonic nonetheless \cite{Xiang}, as shown in Fig.~\ref{Fig: projection probabilities}. \\

\section{Distinguishability transformations for single particle states}  We now turn to single particle states. In this case there is no difference between bosons and fermions, and for the latter this analysis is general and sufficient since, by the Pauli exclusion principle, fermions can only interfere into final states containing one excitation. We show that analogously to the deliberate distinguishability transformation in the two-photon HOM experiment, one can also prepare a single-particle state in an increasingly distinguishable manner in various ways: one way would be to send a single photon into a variable reflectivity, loss-less beam splitter. The larger the probability amplitude for the state corresponding to finding the photon in one of the arms as compared to the other, the more distinguishable the photon paths become. A similar method would be to prepare the photon in a linearly polarized mode. If the photon's polarization is diagonal to the horizontal-vertical (HV) polarization basis, then the two paths expressed in this basis become indistinguishable, but as the polarization of the photon is rotated towards, e.g., the horizontal direction, the paths become increasingly distinguishable in the HV basis.

Such an initial state can be written as \beq \ket{\psi^D_1(\gamma)} = \cos(\pi/4 + \gamma/2) \ket{1,0} + \sin(\pi/4 + \gamma/2) \ket{0,1}. \label{Eq:Initial indistinguishability} \eeq For $\gamma=0$ it is not possible to distinguish or guess in which of the two modes one would ``find'' the particle, if one were to measure, while for $\gamma=\pi/2$ one can be confident, \textit{a priori} of any measurement, to find the particle in the second mode. This state is projected onto the projector state $ \ket{\xi_1} = \cos(\beta_n) \ket{1,0} + e^{-i \theta_n}\sin(\beta_n) \ket{0,1}, \label{Eq:Projector} $ where $\beta_n$ and $\theta_n$ are variables that can be used to implement any pure, two-mode, single-photon projector. Note that here, for simplicity, we have incorporated the beam splitter, or any other unitary transformation, into this projector, and so referring to Fig.~\ref{Fig: schematic}, this projector state represents $\hat{U} \ket{\xi}$. If the two modes are taken to be the horizontally and vertically, linearly polarized modes, the projector can be implemented through a variable birefringence plate implementing $\theta_n$, followed by polarizer oriented with the transmission polarization at the angle $\beta_n$, that in turn is followed by a single-photon detector. The projection probability $P^D_1$ for this input state is
\begin{eqnarray} \fl P^D_1  = \cos^2(\beta_n) \left [1-\sin(\gamma) \right]/2 + \left[\cos(\theta_n)\sin(2\beta_n) \cos(\gamma)\right]/2 \nonumber \\ + \sin^2(\beta_n) \left [1+\sin(\gamma)\right ]/2. \label{Eq:Projector D} \end{eqnarray}

\begin{figure}[ht]
\centering
\includegraphics[scale=0.4]{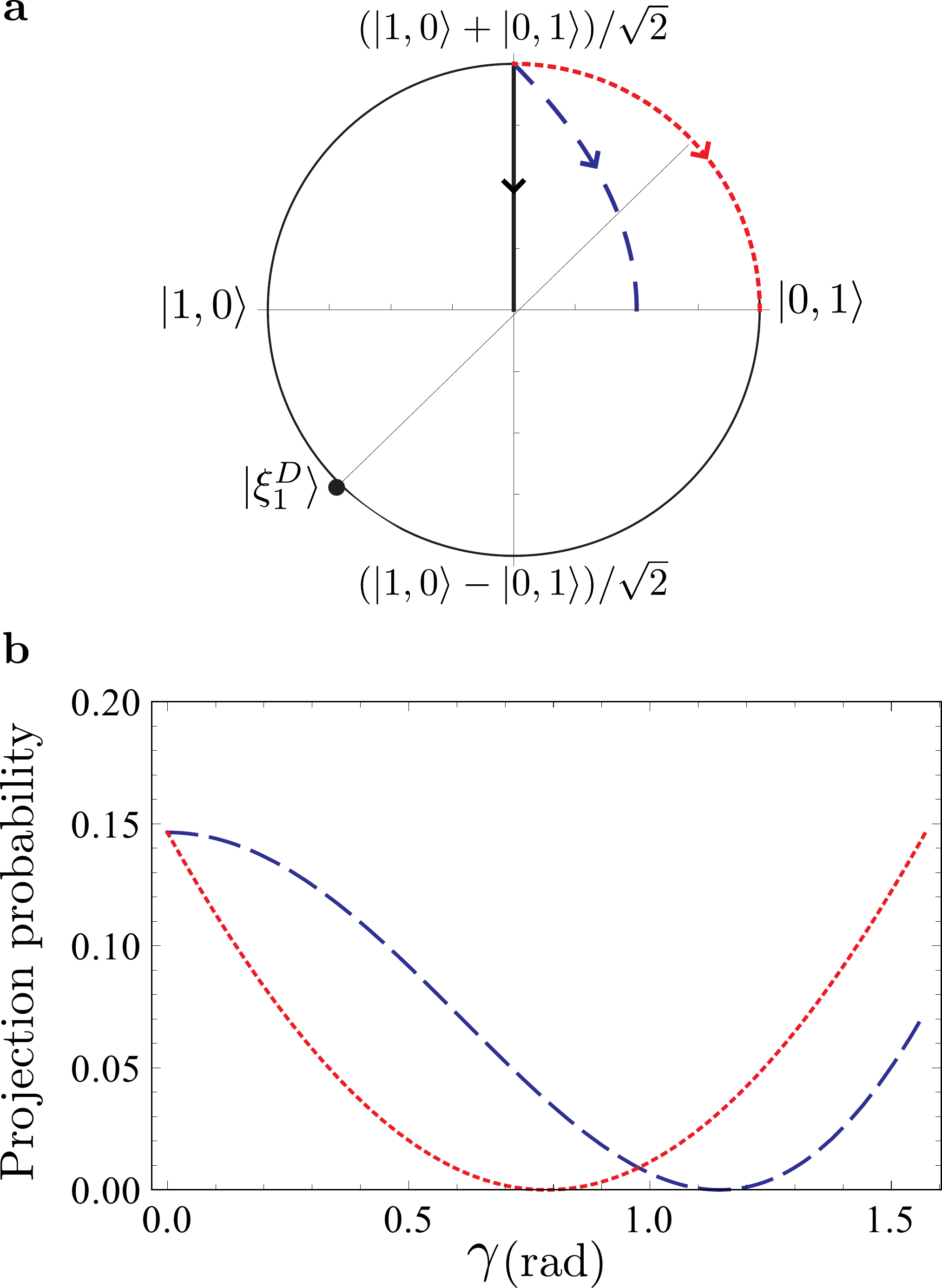}
\caption{\textbf{a.} The Stokes vector evolution for the different distinguishability models. Dotted (red) line indicates a deliberate distinguishability transformation due to a polarization rotation, Eq. (\ref{Eq:Initial indistinguishability}). Dashed (blue) line, indicates the increase of distinguishability due to linear loss, Eq. (\ref{Eq:Initial loss}). Solid (black) line indicates increasing phase noise, Eq. (\ref{Eq:Initial phase noise}). \textbf{b.} Projection probabilities for $P_1^D$ (dotted, red) and $P_1^L$ (dashed, blue) as a function of distinguishability}
\label{Fig: Bloch}
\end{figure}

This projection probability is not monotonic with respect to $\gamma$ for certain values of the other parameters as can be inferred from Fig.~\ref{Fig: Bloch} \textbf{a}. For polarized photons the figure shows the equatorial plane of the Poincar\'{e} sphere. For the projector state $\ket{\xi_1^D}=\cos(\pi/8) \ket{1,0}-\sin(\pi/8)\ket{0,1}$, corresponding to $\beta_n=\pi/8$ and $\theta_n=\pi$ and marked by the black dot in Fig.~\ref{Fig: Bloch} \textbf{a}, the overlap with the partially distinguishable state on the dotted, red line is zero where the thin black line emanating from the black dot and crossing the origin intersects the dotted curve. (The states, being antipodal on the Poincar\'{e} sphere, are orthogonal.) For states with either more or less distinguishability the overlap is finite since the states are non-orthogonal. Hence, the projection probability will be non-monotonic as a function of $\gamma$, with zero as a minimum as can be seen in Fig.~\ref{Fig: Bloch} \textbf{b}.

The projection measurement $\ket{\xi_1^D}\bra{\xi_1^D}$ corresponds to the interference between two spatial modes in an unbalanced beam splitter followed by a photo detector in one of the arms. In a polarization setting it corresponds to a half-wave plate oriented with its axis parallel to the linear polarization basis, followed by a polarizer set at 22.5 degrees and a photo detector. However, if the \emph{proper} distinguishability projector for this basis $\ket{\xi^P_1}=(\ket{0,1} + \ket{1,0})/\sqrt{2}$ is chosen, then the measured probability equals the indistinguishability $I^D_1=\cos^2(\gamma/2)$ that goes from 1 to 0.5 monotonically.

Other models of distinguishability can be analysed in a similar fashion. In the case where one of the mode experiences linear loss, then if the loss is large enough we can be convinced that any detected photon must have resided in ``the other'' mode. In this case, to preserve unitarity and normalization, at least a three mode model is needed. The input state can thus be modeled \beq \ket{\psi^L_1(\gamma)} = \left [\cos(\gamma) \ket{1,0,0} + \ket{0,1,0}+ \sin(\gamma) \ket{0,0,1} \right ]/\sqrt{2}. \label{Eq:Initial loss} \eeq If we assume that any excitation in the rightmost (ancillary) loss mode remains undetected, the projection probability $P^L_1$ onto the single-photon projector $\ket{\xi_1}\bra{\xi_1}$ is
\beq P^L_1 = \left [ \cos^2(\gamma)\cos^2(\beta_n) + \cos(\theta_n)\sin(2\beta_n)\cos(\gamma) +  \sin^2(\beta_n) \right ]/2. \label{Eq:Projector B}\eeq This projection probability is also not monotonic with respect to the parameter $\gamma$ for certain values of the other parameters as can easily be seen from Fig. \ref{Fig: Bloch} \textbf{a} and \textbf{b}, blue, dashed line (which is also drawn for $\beta_n=\pi/8$ and $\theta_n=\pi$). The indistinguishability $I^L_1=[1+\cos(\gamma)]^2/2$, however, is monotonic in the relevant interval $0 \leq \gamma \leq \pi/2$.

As a third model, we shall analyse distinguishability brought about by phase noise, the initial state can be written as,
\beq \ket{\psi^N_1} = \left( \ket{1,0} + e^{i \phi_n} \ket{0,1} \right)/\sqrt{2}, \label{Eq:Initial phase noise} \eeq where $\phi_n$ is a random variable representing the (relative-)phase fluctuations between the modes. The projection probability of this state onto $\ket{\xi_1}\bra{\xi_1}$ is derived to be,
\beq P^N_1 = \left [ 1 +  \langle \cos(\theta_n + \phi_n\rangle \sin(2\beta_n) \right] /2, \label{Eq:Projector with noise}
\eeq
where $\langle \ \rangle$ denotes an ensemble average. If we take $\phi_n$ to have a symmetric distribution around a zero mean, we have $\langle \cos(\theta_n + \phi_n)\rangle = \cos(\theta_n)\langle \cos(\phi_n)\rangle$. In the following we are not interested in modelling any particular physical source of phase noise, but can instead simply use the parametrization $\langle \cos(\phi_n)\rangle = \cos(\gamma)$, $\gamma \in \{0,\pi/2 \}$ to characterize the degree of distinguishability between the two modes due to the fluctuating relative phase. Thus we arrive at our final formula \beq P^N_1 = \left [ 1 + \cos(\theta_n)\sin(2\beta_n)\cos(\gamma) \right ]/2, \label{Eq:Projector A}\eeq From the formula we can see that since $\cos(\gamma)$ is a monotonic function in the interval $\{0,\pi/2 \}$, so is $P^N_1$, irrespective of the value of any of the other parameters. Thus, for this model any single-photon projection probability described by $\ket{\xi_1}$ will either increase or decrease monotonically, or be flat. This can be seen in Fig. \ref{Fig: Bloch} \textbf{a} where the density matrix is transformed along a radial ray in the Poincar\'{e} sphere as a function of increasing phase noise or distinguishability. The indistinguishability in this case is $I^N_1=[1+\cos(\gamma)]/2$ which is monotonic when $0 \leq \gamma \leq \pi/2$.\\

It is well known that when analysing linear interference one can get the classical results by extrapolating the case of single-photon input, i.e., replacing the input probability amplitudes with field amplitudes and interpreting the output probability as a mean output intensity. From this analogy it is clear that if classical fields are used with the mode distinguishability models above, and the corresponding classical interference measurement were done, the corresponding non-monotonic output intensity curves would be obtained. This shows that non-monotonic interference probability does not, in general, stem from a quantum-to-classical transition.

\section{Experiment}
The analysis and the predictions of non-monotonicity for the single-photon states above are so simple that we find it pointless to prove them by experiments. Here we shall therefore discuss the more interesting case of distinguishability (mode) transformation for the two-photon case, experimentally implemented as a polarization rotation. We saw in the analysis of the two-photon Hong-Ou-Mandel experiment that a monotonic behavior followed. This is not the case for all two-photon experiments as we shall show. To this end, two photons prepared in a diagonal, linearly polarized mode will have the maximum indistinguishability with respect to the HV modes. Rotating the diagonally polarized state to become, e.g., horizontally polarized, the distinguishability will become perfect. The initial two mode state can be written as \begin{eqnarray} \fl \ket{\psi^D_2} = \sin^2(\pi/4 + \gamma/2) \ket{2,0} + \sqrt{2} \sin(\pi/4 + \gamma/2)\cos(\pi/4 + \gamma/2) \ket{1,1} \nonumber \\+ \cos^2(\pi/4 + \gamma/2) \ket{0,2}, \label{Eq:Initial two-photon indistinguishability} \end{eqnarray} where $\gamma$ is the relative angle between the input and output polarizations. The two-photon interference projector we shall use for this state is \beq \ket{\xi^D_2} = \frac{1}{\sqrt{3}}\left ( \sqrt{2} \ket{2,0} +  \ket{1,1}  \right ). \label{Eq:Two photon projector D} \eeq It is deliberately chosen to on the one hand involve interference between the two states $\ket{2,0}$ and $\ket{1,1}$, and on the other hand not be the proper indistinguishability projector.
Computing the projection probability, one gets \beq P^D_2  =  \frac{4}{3}  \sin^2(\pi/4 + \gamma/2)\cos^2(\gamma/2). \label{Eq: prediction}\eeq
For a diagonally polarized input state, $\gamma=0$, one gets $P^D_2=2/3$.
\begin{figure}[ht]
\centering
\includegraphics[scale=0.4]{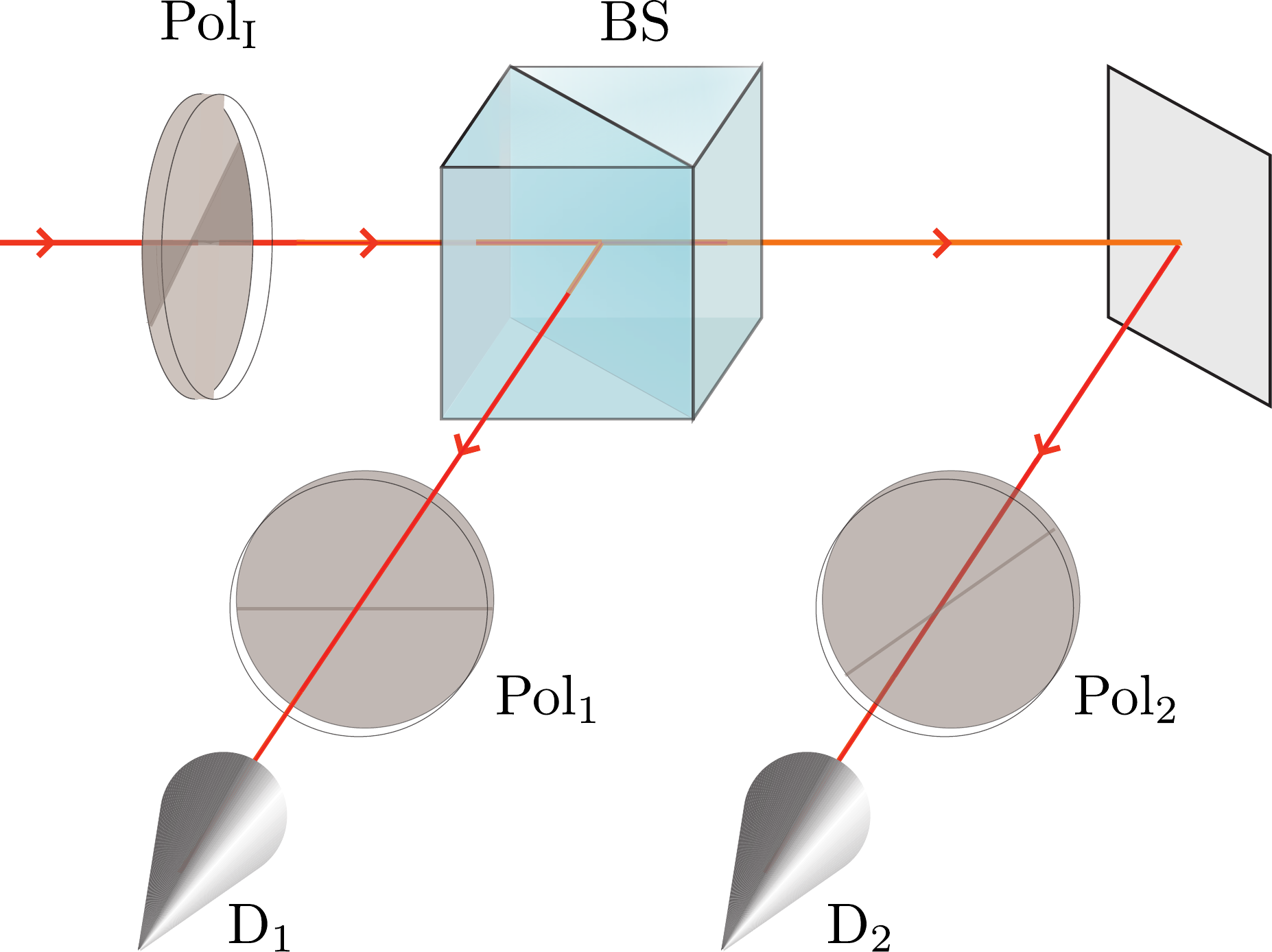}
\caption{Experimental setup. Circularly polarized light falls on a polarizer (Pol$_{\scriptsize \textrm{I}}$) oriented at 45 degrees initially. The polarization is rotated from diagonal to horizontal (depicted by the grey shaded region). A non-polarizing beam splitter (BS) splits the beam into two arms; one with a polarizer with transmission axis oriented horizontally (Pol$_{\scriptsize \textrm{1}}$) and the other polarizer oriented diagonally (Pol$_{\scriptsize \textrm{2}}$). Coincidence between the two detectors (D$_{\scriptsize \textrm{1}}$ and D$_{\scriptsize \textrm{2}}$) then projects out the required two-photon state.}
\label{Fig: setup}
\end{figure}

\begin{figure}[h!]
\centering
\includegraphics[scale=0.5]{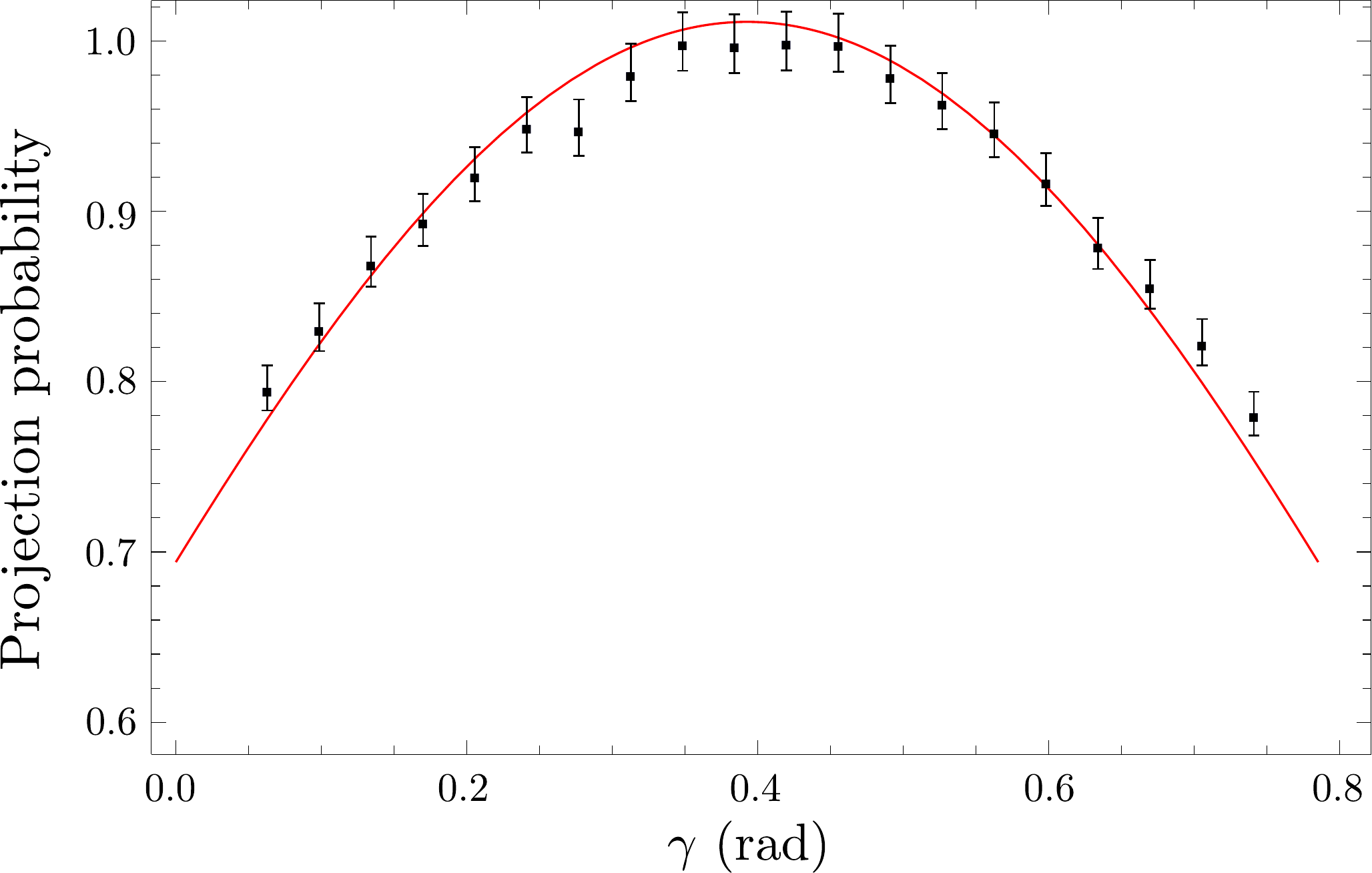}
\caption{Two-photon distinguishability polarization transformation. The black squares represent experimental data and the red curve is $A P^D_2$ with $A$ chosen for best fit. The data is normalized with respect to the maximum number of counts. Error bars shows $\pm \sigma$.}
\label{Fig: exp}
\end{figure}

In \cite{Hofmann} it was shown that any $N$ photon, two-mode projector can be implemented probabilistically by writing the state, expressed in terms of creation operators of the two modes, as a product of single-photon creation operators. For the state $\xi^D_2$ above the decomposition becomes
\beq \xi^D_2  = \sqrt{\frac{2}{3}} \frac{\left ( \adag + \bdag  \right ) }{\sqrt{2}}  \adag \ket{0,0}
\label{Eq: Implementation D} \eeq
The consequence is that the projector can probabilistically be implemented by splitting the diagonally polarized, two-photon state in a non-polarizing 50:50 beam splitter, there is a 50 \% chance that one photon exits from each beam splitter output port. Then after one port a polarizer is inserted that transmits diagonally polarized photons followed by a photo detector. Each time a diagonally polarized photon exits this port an ideal detector will click. After the second port one inserts a polarizer that transmits horizontally polarized photons. Hence, if one diagonally polarized photon exits this port there is a 50 \% chance that an ideal detector clicks. If one looks at the probability $P^E_2$ that the two detectors click in coincidence one finds that (for ideal detectors) the probability is $1/4$. Hofmann proved that $P^E_2 \propto P^D_2$ \cite{Hofmann}. Comparing the two expressions for a diagonally polarized, two-photon, input state we got 1/4 and 2/3, respectively. Hence, the coincidence probability $P_2^E$ will be connected to the projection probability $P_2^D$ as
\beq
P^E_2 = \frac{3}{8} P^D_2.\eeq Taking non-unity photo detection into account one obtains $P^E_2 = 3 \eta^2 P^D_2/8$, where it has been assumed that the detector efficiency $\eta$ is the same for both detectors.
Fig.~\ref{Fig: exp} shows the experimental results - implemented via post-selection of the two-photon component of a weakly excited coherent state input \cite{Shabbir} - along with the theoretical curve. As can be seen, and as predicted by (\ref{Eq: prediction}), the curve is non-monotonic.

Using classical light, with field amplitude $E_0$, the output of the above experiment would be $(E_0/2)^4 \cos^2(\gamma-\theta_1)\cos^2(\gamma-\theta_1)$, with $\theta_1$ and $\theta_2$ as 0 and $\pi/4$ respectively. This is also a non-monotonic function of the input polarization $\gamma$.

\section{Discussion} In this paper we have discussed various ways by which states can be made increasingly distinguishable with respect to a given measurement basis and we have shown that by suitable choice of interference projectors, non-monotonic projection probabilities can be demonstrated for most states, including single particle states and classical states detected either classically or with single particle detectors. Hence, in general non-monotonicity can neither be regarded as a consequence of \emph{quantum} interference between multiple photons nor as a quantum-to-classical transition phenomenon \cite{Ra, Tichy}. However, given an initial state in a particular measurement basis, a suitable ``maximally indistinguishable'' projector can be defined. We have named such a projector the \textit{proper} indistinguishability projector. With respect to this choice, the interference projection probability always change monotonically as a function of increasing distinguishability. Thus we see that the choice of measurement projector critically determines how the measurement probability behaves as a function of the distinguishability between the interfering modes, particles, or states.

\ack We gratefully acknowledge discussions with Prof. Y.-H. Kim and Mr. Y.-S. Ra. This work was supported by the Swedish Research Council (VR) through grant 621-2011-4575 and through its support of the Linn\ae us Excellence Center ADOPT.


\section*{References}
\begin{thebibliography}{99}

\bibitem{HBT} R. Hanbury Brown and R. Q. Twiss, Nature \textbf{178}, 1046 (1956). 

\bibitem{HBT2} R. Hanbury Brown and R. Q. Twiss, Proc. Royal Soc. London A \textbf{243}, 291 (1958). 

\bibitem{HOM} C. K. Hong, Z. Y. Ou, and L. Mandel, Phys. Rev. Lett. \textbf{59} 18 (1987).

\bibitem{Ra} Y.-S. Ra, M. C. Tichy, H.-T. Lim, O. Kwon, F. Mintert, A. Buchleitner, and Y.-H. Kim, Proc. Natl. Acad. Sci. USA \textbf{110} 1227 (2013).

\bibitem{Tichy} M. C. Tichy, H.-T. Lim, Y.-S. Ra, F. Mintert, Y.-H. Kim, and A. Buchleitner, Phys. Rev. A \textbf{83}, 062111 (2011).

\bibitem{Jaeger} G. Jaeger, M. A. Horne, and A. Shimony, Phys. Rev. A \textbf{48}, 1023 (1993).

\bibitem{Englert} B.-G. Englert, Phys. Rev. Lett. \textbf{77}, 2154 (1996).

\bibitem{Durr} S. Durr, T. Nonn, and G. Rempe, Phys. Rev. Lett. \textbf{81}, 5705 (1998).

\bibitem{Bjork} G. Bj\"{o}rk and A. Karlsson, Phys. Rev. A, \textbf{58}, 3477 (1998).

\bibitem{Gavenda} M. Gavenda, L. \u{C}elechovsk\'{a}, J. Sobusta, M. Du\u{s}ek, and R. Filip, Phys. Rev. A \textbf{83}, 042320 (2011).

\bibitem{Xiang} G.Y. Xiang, Y. F. Huang, F. W. Sun, P. Zhan, Z. Y. Ou, and G. C. Guo, Phys. Rev. Lett. \textbf{97}, 023604 (2006).

\bibitem{Hofmann} H. Hofmann, Phys. Rev. A \textbf{70}, 023812 (2004).

\bibitem{Shabbir} S. Shabbir, M. Swillo, G Bj\"ork, Phys. Rev. A \textbf{87}, 053821 (2013).

\end{thebibliography}
\end{document}